\documentclass[preprint,secnumarabic,amssymb, nobibnotes, aps, prl,superscriptaddress]{revtex4-1}

\usepackage{epsfig}
\usepackage{subfigure}
\usepackage{amsmath,amssymb,color}
\usepackage[english]{babel}
\usepackage{hyperref}
\usepackage{setspace}

\usepackage[capitalize]{cleveref}

\newcommand{\be}{\begin{equation}}
\newcommand{\ee}{\end{equation}}

\newcommand{\barr}{\begin{array}}
\newcommand{\earr}{\end{array}}

\newcommand{\beqn}{\begin{eqnarray}}
\newcommand{\eeqn}{\end{eqnarray}}

\newcommand{\bs}{\begin{subequations}}
\newcommand{\es}{\end{subequations}}

\newcommand{\bw}{\begin{widetext}}
\newcommand{\ew}{\end{widetext}}

\newcommand{\f}{\frac}

\newcommand{\op}[1]{\operatorname{#1}}

\begin{document}

\title{Multicritical Scaling in a Lattice Model of Vesicles}

\author{N. Haug}
\affiliation{School of Mathematical Sciences, Queen Mary University of London, London, E1 4NS, United Kingdom}
\affiliation{Section for Science of Complex Systems, CeMSIIS, Medical University of Vienna, Spitalgasse 23, A-1090 Vienna, Austria}
\affiliation{Complexity Science Hub Vienna, Josefst\"adter Stra\ss e 39, A-1080 Vienna, Austria}
\author{T. Prellberg}

\affiliation{School of Mathematical Sciences, Queen Mary University of London, London, E1 4NS, United Kingdom}
\date{\today}

\begin{abstract}
Vesicles, or closed fluctuating membranes, have been modeled in two dimensions by self-avoiding polygons, weighted with respect to their perimeter and enclosed area, with the simplest model given by area-weighted excursions. These models generically show a tricritical phase transition between an inflated and a crumpled phase, with a scaling function given by the logarithmic derivative of the Airy function. 
Extending such a model, we find realizations of multicritical points of arbitrary order, with the associated multivariate scaling functions expressible in terms of generalized Airy integrals, as previously conjectured by John Cardy.
This work therefore adds to the small list of models with a critical phase transition, for which exponents and the associated scaling functions are explicitly known.

\end{abstract}

\maketitle

\indent{\it Introduction.}---Obtaining a thorough understanding of phase transitions is one of the main aims of statistical physics. For  a continuous transition one would like to know the critical exponents describing the singular power-law behaviour of thermodynamic quantities as the transition is approached. Moreover, in the vicinity of such a transition it is generally believed that the thermodynamic quantities depend only on a suitably scaled combination of the parameters in terms of a universal scaling function \cite{Cardy96}. Most progress has been made in two dimensions with the help of conformal invariance \cite{Henkel2013}. While these scaling functions can be easily obtained numerically, there are only few instances for which one knows precise expressions, one classical example being the spin-spin correlations of the two-dimensional Ising model \cite{Wu76}. Based on field-theoretic arguments, John Cardy postulated that by including many-body interactions in a model of vesicles, a hierarchy of scaling functions could be found, expressed in terms of
generalized, higher-order Airy integrals
\begin{equation} 
\Theta_\ell(s_1,s_2,\dots,s_\ell)= \f{1}{2\pi i}\int_{C} \exp\left(\f{u^{\ell+2}}{\ell+2}-\sum_{j=1}^{\ell}s_{j} u^{j}\right)du,
\label{eq_generalized_Airy}
\end{equation}
with a suitable contour $C$. However, he cautioned that due to the technical limitations of the method used, ``it is very difficult to say to what these higher multicritical points might correspond physically'' \cite{Cardy01}. In this letter, we give explicit examples of a statistical mechanical model having precisely these scaling functions, thereby providing a resolution to this problem.

\indent{\it Vesicles and Self-Avoiding Polygons.}---A vesicle consists of a closed membrane formed from a lipid bilayer inside a watery solution. Depending on parameters such as the temperature and the osmotic pressure difference between the outside and the inside of the membrane, vesicles are found in different typical conformations \cite{Seifert97}. 
Subject to thermal fluctuations, a vesicle of fixed surface area favours ``crumpled'' configurations with relatively small volume if there is a large net pressure acting onto the outside of the membrane. On the other hand, if there is a net pressure acting onto the inside of the membrane, then the vesicle tends to appear in an inflated shape with larger volume.

\begin{figure}[h]
\begin{center}
\includegraphics[width=0.25\textwidth,angle=0]{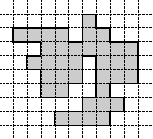}
\end{center}
\caption{A self-avoiding polygon (SAP) on $\mathbb{Z}^2$ of perimeter 52 and area 37.}
\label{fig_sap}
\end{figure}

In \cite{Leibler89}, a two-dimensional model of vesicles was proposed in terms of ring polymers enclosing an area. In this case, the length of the polymer plays the role of the surface area of the vesicle, and the volume of the vesicle becomes the enclosed area. In \cite{Banavar91,Fisher91} the vesicles were modeled as self-avoiding polygons (SAP) on the square lattice -- see \cref{fig_sap} for an example. Note that any intrinsic property of the vesicle membrane such as stiffness is neglected in that lattice model.

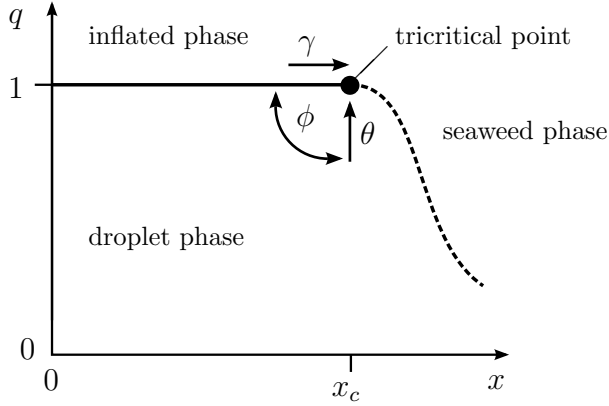
\begin{figure}[htb]
  \centering
  \def\svgwidth{200pt}
\begingroup
  \makeatletter
  \providecommand\color[2][]{
    \errmessage{(Inkscape) Color is used for the text in Inkscape, but the package 'color.sty' is not loaded}
    \renewcommand\color[2][]{}
  }
  \providecommand\transparent[1]{
    \errmessage{(Inkscape) Transparency is used (non-zero) for the text in Inkscape, but the package 'transparent.sty' is not loaded}
    \renewcommand\transparent[1]{}
  }
  \providecommand\rotatebox[2]{#2}
  \ifx\svgwidth\undefined
    \setlength{\unitlength}{7.5cm}
    \ifx\svgscale\undefined
      \relax
    \else
      \setlength{\unitlength}{\unitlength * \real{\svgscale}}
    \fi
  \else
    \setlength{\unitlength}{\svgwidth}
  \fi
  \global\let\svgwidth\undefined
  \global\let\svgscale\undefined
  \makeatother
  \begin{picture}(1,0.76172539)
    \put(0,0){\includegraphics[width=\unitlength]{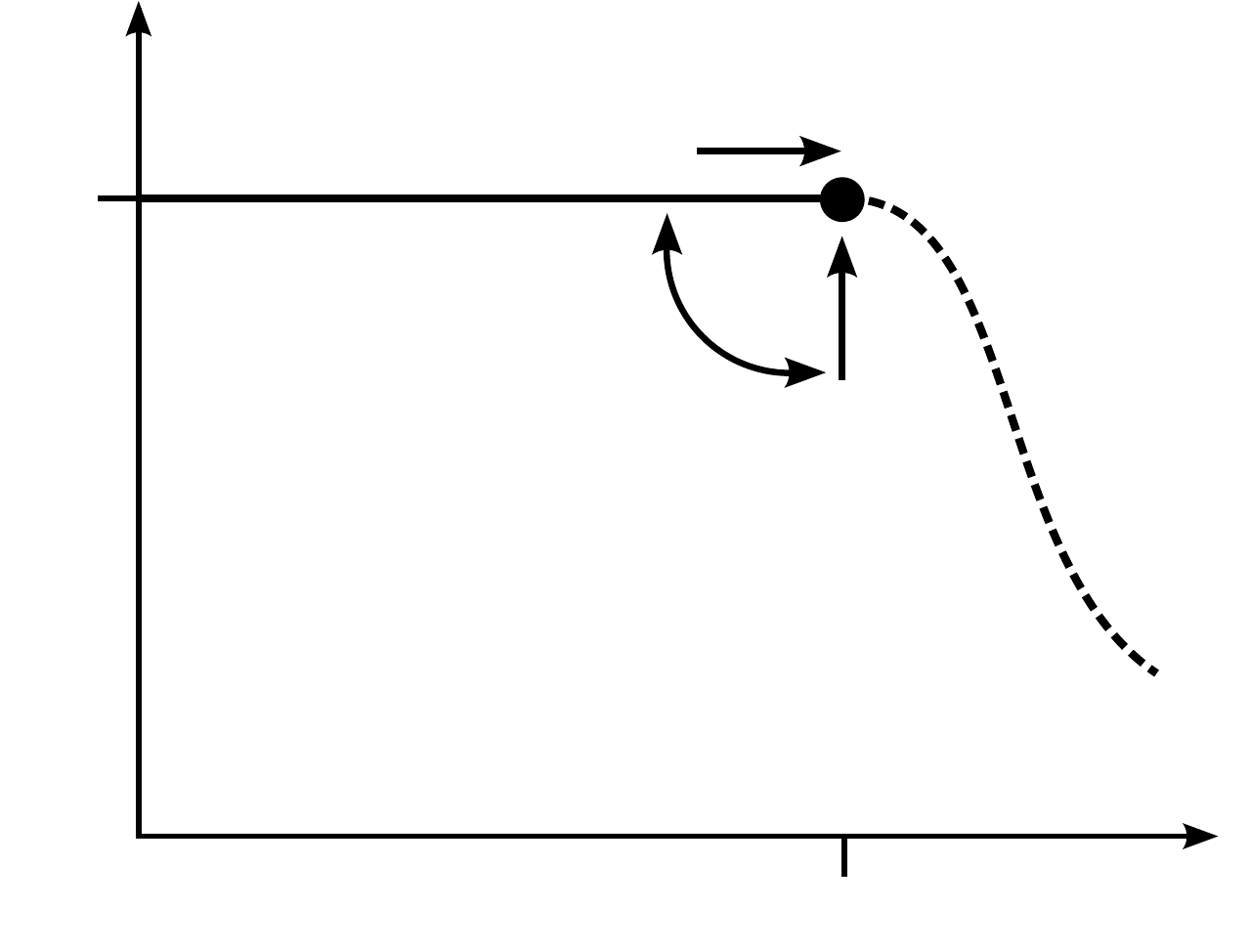}}
    \put(0.02802149,0.61607872){\color[rgb]{0,0,0}\makebox(0,0)[lt]{\begin{minipage}{0.94191589\unitlength}\raggedright $1$\end{minipage}}}
    \put(0.028937,0.74495898){\color[rgb]{0,0,0}\makebox(0,0)[lt]{\begin{minipage}{0.19035234\unitlength}\raggedright $q$\end{minipage}}}
    \put(0.64040718,0.0389639){\color[rgb]{0,0,0}\makebox(0,0)[lt]{\begin{minipage}{1.06334752\unitlength}\raggedright $x_c$\end{minipage}}}
    \put(0.93477791,0.02676937){\color[rgb]{0,0,0}\makebox(0,0)[lb]{\smash{$x$}}}
    \put(0.0950392,0.06475949){\color[rgb]{0,0,0}\makebox(0,0)[lt]{\begin{minipage}{0.38398662\unitlength}\raggedright $0$\end{minipage}}}
    \put(0.05,0.12262628){\color[rgb]{0,0,0}\makebox(0,0)[lt]{\begin{minipage}{0.38398662\unitlength}\raggedright $0$\end{minipage}}}
    \put(0.69371364,0.53209235){\color[rgb]{0,0,0}\makebox(0,0)[lt]{\begin{minipage}{1.02724626\unitlength}\raggedright $\theta$\end{minipage}}}
    \put(0.57673978,0.69373077){\color[rgb]{0,0,0}\makebox(0,0)[lt]{\begin{minipage}{0.94847975\unitlength}\raggedright $\gamma$\end{minipage}}}
    \put(0.57279818,0.56723095){\color[rgb]{0,0,0}\makebox(0,0)[lt]{\begin{minipage}{0.3043404\unitlength}\raggedright $\phi$\end{minipage}}}
    \put(0.67,0.6){\line(1,1){0.075}}
    \put(0.76,0.68){\footnotesize tricritical point}
    \put(0.18,0.3){\footnotesize droplet phase}
    \put(0.18,0.68){\footnotesize inflated phase}
    \put(0.85,0.5){\footnotesize seaweed phase}    
  \end{picture}
\endgroup

   \caption{Qualitative picture of the phase diagram of the SAP model of vesicles. The line $q_c(x)$ is the boundary of the droplet phase. The exponents $\gamma,\theta$ and $\phi$ characterize the singular behaviour of the generating function $P(x,q)$ around the tricritical point.}

  \label{fig_phasediag}
\end{figure}

In order to analyze the model of SAP, one defines the area-perimeter generating function
\begin{equation}
P(x,q)= \sum_{m=0}^\infty \sum_{n=0}^\infty p_{m,n} x^m q^n,
\label{eq_genfun_SAP}
\end{equation}
where $p_{m,n}$ is the number of SAP with perimeter $m$ and area $n$, with two SAP being considered identical if they are the same up to translation.
The qualitative behaviour of the radius of convergence $q_c(x)$ of $G(x,q)$, seen as a series in $q$ for fixed values of $x$, was discussed in \cite{Fisher91,Banavar91}. This quantity is closely related to the asymptotic growth rate of the partition function $Z_n(x)=\sum_{m=0}^\infty p_{m,n} x^m$, and thus physically to the free energy per unit area in the thermodynamic limit of infinite area. This model exhibits a phase transition at a value $x_c$ at which $q_c(x)$ is not analytic. More precisely, it was shown that there exists a value $x_c>0$ such that for $0\leq x\leq x_c$, $q_c(x) = 1$; for $x>x_c$, $q_c(x)$ is a continuous function of $x$, $q_c(x)<1$ and $\lim_{x\to \infty} q_c(x) = 0$ -- see \cref{fig_phasediag}. For $q<q_c(x)$, polygons with relatively small area dominate the sum \eqref{eq_genfun_SAP}. This part of the $(x,q)$-plane is  called the droplet phase. Analogously, the region $q>1$ is labeled the inflated phase. The region where $q_c(x)<q<1$ is described as the `seaweed' phase, in which the typical conformation consists of a space-filling, convoluted polygon. Exact enumerations yield the estimate $x_c\simeq 0.379$ \cite{Jensen99}. The point $(x,q)=(x_c,1)$ is called a tricritical point \cite{Lawrie84}.

In \cite{Richard01,Richard04}, exact enumeration data was used to analyse the singular behaviour of the area-perimeter generating function of \emph{rooted} SAP. In rooted SAP, there exists one distinguished point on the perimeter of the SAP, therefore the number of rooted SAP with perimeter $m$ and area $n$ is $m p_{m,n}$, and the area-perimeter generating function is $R(x,q)=x\f{d}{dx}P(x,q)$. It was conjectured that in the vicinity of the point $(x,q)=(x_c,1)$, the singular part of this function satisfies the scaling relation 
\begin{equation}
R^{\op{sing}}\left(x,e^{-\epsilon}\right) = \epsilon^{\theta} F((x_c-x)\epsilon^{-\phi}),
\label{eq_scalingrel_RP}
\end{equation}
with the scaling function being, up to prefactors, given by the logarithmic derivative of the \href{https://dlmf.nist.gov/9}{Airy function}, which is defined for $z\in \mathbb{C}$ as \cite{NIST} %
\begin{equation}
\op{Ai}(z) = \f{1}{2\pi i} \int_{\infty e^{-i\pi/3}}^{\infty e^{i\pi/3}} \exp\left(\f{u^3}{3}-z u\right) du.
\end{equation}

In \cite{Cardy01} it was argued via field theoretic methods that, upon introducing further interactions into the SAP model, one should be able to observe multicritical points of higher order, described by scaling functions of more than one variable. More precisely, upon introducing $\ell-1$ further interactions $(w_j)_{j=1}^{\ell-1}$, there should exist multicritical points in the vicinity of which the singular part of the multivariate generating function of rooted SAP satisfies the scaling relation
\begin{equation}
R^{sing}(w_1,\dots,w_{\ell-1},x,q) = \epsilon^{\theta}F(\alpha_1\epsilon^{\phi_1},\alpha_2\epsilon^{\phi_2},\dots,\alpha_\ell ^{\phi_\ell}),
\label{eq_scaling_SAP}
\end{equation}
where the variables $(\alpha_j)_{j=1}^\ell$ depend on the parameters of the generating function, the crossover exponents $\phi_j$ are given by
\begin{equation}
\phi_j=\phi_j(\ell)=\f{\ell+2-j}{\ell+2}\quad(1\leq j \leq \ell),
\label{eq_crit_exps}
\end{equation}
and 
$\theta=\f{1}{\ell+2}$ is a critical exponent. The scaling function $F(s_1,s_2,\dots,s_\ell)$ is expressible in terms of generalized, higher-order Airy integrals, defined in Eq. (\ref{eq_generalized_Airy}).
However, no details of the interactions necessary to observe these multicritical points were given in that reference.

\indent{\it The Model.}---One-dimensional lattice paths occur in many applications in probability theory, combinatorics and statistical physics. For $
m\in \mathbb{N}_0 = \mathbb{N}\cup\{0\}$, a one-dimensional lattice path of length $m$ is a sequence $(\mathbf{r}_0,\mathbf{r}_2,\dots,\mathbf{r}_m)$ of points of $\mathbb{Z}^2$, where for $0<j\leq m$, $\mathbf{r}_j-\mathbf{r}_{j-1}\in \{1\} \times \mathcal{S}$, with $\mathcal{S}\subseteq \mathbb{Z}$ \cite{Banderier02}. One usually fixes $\mathbf{r}_0=(0,0)$. The path then stays in the right half-plane. Paths restricted further to stay in the upper right quarter plane $\mathbb{N}_0\times\mathbb{N}_0$ are called meanders, paths which end on the horizontal line $\mathbb{N}_0 \times \{0\}$ are called bridges, and paths which are both meanders and bridges are called excursions. \L{}ukasiewicz paths, which encode rooted ordered trees \cite{LeGall05}, are excursions with $\mathcal{S}=\{k \in \mathbb{Z}~|~k\geq -1\}$. Excursions with $\mathcal{S}=\{-1\}\,\cup\,\{j~|~k \leq j \leq \ell\}$, where $k,\ell\in \mathbb{N}_0$, are called $(k,\ell)$-\L{}ukasiewicz paths \cite{Brak11}.
\begin{figure}[h]
\begingroup
  \makeatletter
  \providecommand\color[2][]{
    \errmessage{(Inkscape) Color is used for the text in Inkscape, but the package 'color.sty' is not loaded}
    \renewcommand\color[2][]{}
  }
  \providecommand\transparent[1]{
    \errmessage{(Inkscape) Transparency is used (non-zero) for the text in Inkscape, but the package 'transparent.sty' is not loaded}
    \renewcommand\transparent[1]{}
  }
  \providecommand\rotatebox[2]{#2}
  \ifx\svgwidth\undefined
    \setlength{\unitlength}{7.5cm}
    \ifx\svgscale\undefined
      \relax
    \else
      \setlength{\unitlength}{\unitlength * \real{\svgscale}}
    \fi
  \else
    \setlength{\unitlength}{\svgwidth}
  \fi
  \global\let\svgwidth\undefined
  \global\let\svgscale\undefined
  \makeatother
  \begin{picture}(1,0.33736701)
    \put(0,0){\includegraphics[width=\unitlength]{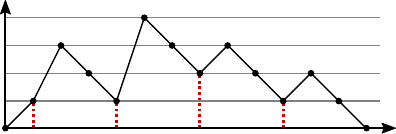}}
    \put(-0.05,0.3){$y$}
    \put(1,-0.025){$x$}
  \end{picture}
\endgroup
\caption{A $(1,3)$-\L{}ukasiewicz path of length 13.}
\label{fig_Lukasciewiczpath}
\end{figure}
 \Cref{fig_Lukasciewiczpath} shows a $(1,3)$-\L{}ukasiewicz path of length 13, with the horizontal and vertical axes of $\mathbb{Z}^2$ labeled by $x$ and $y$, respectively. A step in the direction $(1,-1)$ is called a down-step and a step in direction $(1,k)$, where $k\geq 1$ is called an up-step of length $k$. The \emph{height} of a point is its distance from the $x$-axis, and the height of a step is the height of its initial point. The initial points of the up-steps in \cref{fig_Lukasciewiczpath} are marked by red dotted lines. Special subclasses of \L{}ukasiewicz paths are Dyck and Motzkin paths, corresponding to $(k,\ell) = (1,1)$ and $(k,\ell)=(0,1)$, respectively. 

In this letter we consider the generating function
\begin{equation}
G_{\ell}(\mathbf{w},x,q) = \sum_{m,j_1,\ldots,j_\ell} c(j_1,\dots,j_\ell,m,n) w_1^{j_1} \dots w_\ell^{j_\ell} x^m q^n,
\end{equation}
where we abbreviate $w_1,\dots,w_\ell=\mathbf{w}$ and where $c(j_1,\dots,j_\ell,m,n)$ is the number of $(1,\ell)$-\L{}ukasiewicz paths with $m$ up-steps, of which $j_1$ have length one, $j_2$ have length two etc., and the sum of the heights of all the up-steps is $n$, which is an area-like quantity (in the case of Dyck paths, the number of lattice points below the path is precisely $2n+m$). For example, the path shown in \cref{fig_Lukasciewiczpath} has the weight $w_1^3 w_2 w_3 x^5 q^5$ in the generating function $G_3(w_1,w_2,w_3,x,q)$. 
One can set $w_1=1$ without loss of generality. In the following we therefore write $G_\ell(1,w_2,\dots,w_\ell,x,q)\equiv G_\ell(w_2,\dots,w_\ell,x,q)$.

\indent{\it Previous Results.}---In \cite{Haug15}, the asymptotic behaviour of $G_1(x,q)$ in the limit $q\to 1^-$ was analysed, despite the parameters being interpreted slightly differently there. 
In particular it was shown that, in the vicinity of the tricritical point $(x,q)=(\f14,1)$, the singular part of the generating function satisfies the scaling relation
\begin{equation}
G_1^{\op{sing}}\left(x,e^{-\epsilon}\right) = \epsilon^{\theta} F((x_c-x)\epsilon^{-\phi}),
\label{eq_scalingrel_DP}
\end{equation}
where $\theta=\f13$, $\phi=\f23$ and $F(s)=b_0 \f{d}{ds}\ln(\op{Ai}(b_1s))$, with positive constants $b_0$ and $b_1$.
Consistently with the solution $G_1(x,1)=\f{1}{2x}(1-\sqrt{1-4x})$,  \cref{eq_scalingrel_DP} implies with the asymptotic expansions of the Airy function and its derivative \cite{NIST} that $G_1^{\op{sing}}(x,1)\sim 2(x_c-x)^\gamma$, where $\gamma=\f{\theta}{\phi}=\f12$.
Up to different constants $b_0$ and $b_1$, the scaling relation \eqref{eq_scalingrel_DP} is identical to the one in Eq. (\ref{eq_scalingrel_RP}) that was conjectured to hold for rooted self-avoiding polygons.

\indent{\it Result.}---In \cite{Haug17}, it was shown rigorously that around the multicritical point $(w_2,x,q)=(-\f19,\f13,1)$, the singular part of $G_2(w_2,x,q)$ satisfies the scaling relation
\begin{equation}
G^{\op{sing}}_2(w_2,x,q) = \epsilon^{\theta} F(\alpha_1\epsilon^{-\phi_1},\alpha_2\epsilon^{-\phi_2}),
\end{equation}
where the scaling variables $\alpha_1$ and $\alpha_2$ are analytic functions of $w_2$ and $x$, $\theta=\f14,\phi_1=\f34, \phi_2=\f12$, and the scaling function $F$ is expressible via $\Theta_2(s_1,s_2)$.

This result is generalized in this letter. More specifically, we show that for arbitrary $\ell\geq 2$, there exists a multicritical point $(w_2,\dots,w_\ell,x,q)$ in the model of $(1,\ell)$-\L{}ukasiewicz paths, with $x=(\ell+1)^{-1}, q=1$ and 
\begin{equation} 
w_k = \f{2}{\ell(\ell+1)} \left(\f{-1}{\ell+1}\right)^{k-1} {\ell+1 \choose k+1}\quad(2\leq k \leq \ell), 
\label{eq_critical_jump_weights}
\end{equation}
in the vicinity of which the generating function $G_{\ell}(w_2,\dots,w_\ell,x,q)$ satisfies a scaling relation of the form  of \cref{eq_scaling_SAP} with the same scaling function and the same critical exponents as predicted in \cite{Cardy01}. We thus present an exactly solvable model representing a concrete realization of the multicritical scaling postulated in that reference.

\indent{\it Method.}---To obtain an asymptotic expression for $G_\ell(w_2,\dots,w_\ell,x,q)$ in the limit $q\to 1^-$ for arbitrary $\ell\geq 2$, one proceeds analogously to \cite{Haug15,Haug17}. 
From a simple factorization argument \cite{Brak11} one obtains the functional equation
\begin{equation} 
G_\ell(\mathbf{w},x,q) = 1 + x G_\ell(\mathbf{w},x,q) \left(\sum_{k=1}^\ell w_k \prod_{j=1}^{k} G_\ell(\mathbf{w},q^jx,q)\right).
\label{eq_fun_eq_latticepath}
\end{equation}
We linearize \cref{eq_fun_eq_latticepath} by using the ansatz
\begin{equation} 
G_\ell(w_2,\dots,w_\ell,x,q) = \f{\Phi(qx)}{\Phi(x)},
\label{eq_linearization_ansatz}
\end{equation}
where $\Phi(x)\equiv \Phi(w_2,\dots,w_\ell,x,q)$, 
The solution of the linearized equation is then given by the $q$-hypergeometric series \cite{Gasper90}
\begin{equation} 
\Phi(x) = \sum_{n=0}^\infty \f{\prod_{j=1}^{\ell-1}(\omega_j;q)_n}{(q;q)_n} (-x)^n q^{n^2-n},
\end{equation}
where $(z;q)_n=\prod_{j=0}^{n-1} (1-zq^j)$ is the $q$-Pochhammer symbol and the parameters $(\omega_k)_{k=1}^{\ell-1}$ satisfy%
\begin{equation}
w_k = \f{(-1)^k}{k!} \sum_{j_1=1}^{\ell-1} \sum_{j_2=1}^{\ell-1} \cdots \sum_{j_{k-1} = 1}^{\ell-1} \prod_{p=1}^{k-1} \omega_{j_p}\quad(1\leq k \leq \ell).
\end{equation}
Using the identity
\begin{equation} 
\f{(-1)^{n+1}q^{{n\choose 2}}}{(q;q)_n(q;q)_\infty}=\op{Res}\left[(z;q)_\infty^{-1};z=q^{-n}\right],
\end{equation}
we obtain for $k\in \mathbb{Z}$ the integral expression
\begin{equation} 
\Phi(q^k x) = \f{A}{2\pi i}\int_{C}  \f{z^{\f12(\log_q(z)+1)-\log_q(x)}}{z^k\left(\prod_{j=1}^{\ell-1} (\omega_j/z;q)_\infty\right)(z;q)_{\infty}}dz,
\label{eq_phi_limit_formula}
\end{equation}
where the prefactor $A$ is independent of $k$, and $C$ is a suitably chosen complex contour. Substituting an asymptotic expression for the $q$-Pochhammer symbol \cite{Prellberg95}, the above integral satisfies
\begin{equation} 
\Phi(q^kx) \sim \f{A}{2\pi i}\int_C \exp\left(\f{1}{\epsilon}f(z)\right)\f{g(z)}{z^k}dz\quad(q=e^{-\epsilon} \to 1^-),
\label{eq_asymptotic_integral_expr}
\end{equation}
where the functions are
\begin{align}
f(z)&= \log(x)\log(z)-\f{\log(z)^2}{2} +\sum_{j=1}^{\ell-1}\op{Li}_2\left(\f{\omega_j}{z}\right), \\
g(z)&= \left(\f{z^{\ell}}{(1-z)\prod_{j=1}^{\ell-1}(z-\omega_j)}\right)^{\f12},
\end{align}
and where $\op{Li}_2(z)$ is the \href{http://dlmf.nist.gov/25.12}{Euler dilogarithm} \cite{NIST}. The saddle points of $f(z)$ are the zeros of the polynomial
\begin{align}
\chi(z) &= x \left(\prod_{j=1}^{\ell-1} z-\omega_j\right) - z^{\ell}(1-z) \\ &= z^{\ell+1}\left(1-\f1z + \f{x}{z}\sum_{j=1}^{\ell} \f{w_j}{z^j}\right)\quad (z \neq 0).\nonumber
\end{align}
Comparing the bracket in the last expression with \cref{eq_fun_eq_latticepath}, we see that the solution of \cref{eq_fun_eq_latticepath} for $q=1$ is equal to the inverse of a saddle point of $f(z)$. If we set the weights for $2\leq k \leq \ell$ to the ones given in \cref{eq_critical_jump_weights}, 
then $\ell+1$ saddle points coalesce in the point $z=(\ell+1)^{-1}$ for $x=(\ell+1)^{-1}$. To obtain an asympotic expression for $\Phi(q^k x)$, we apply a method devised in \cite{Ursell72}, based on a theorem from \cite{Levinson61}, from which it follows that if the parameters of the function $f(z)$ are close to the critical values given in \cref{eq_critical_jump_weights}, then there exists a mapping $\op{T}:u\mapsto z(u)$, which is analytic and bijective in the vicinity of the point $z=(\ell+1)^{-1}$, such that
\begin{equation}
f(z(u)) = \f{u^{\ell+2}}{\ell+2}-\sum_{j=0}^\ell \alpha_j u^j = p(u).
\end{equation}
Moreover, the coefficients $(\alpha_j)_{j=0}^\ell$ are analytic functions of $(w_j)_{j=2}^{\ell}$ and $x$ in the region around the point of coalescence of the saddle points. Using the transformation $\op{T}$, the integral in \cref{eq_asymptotic_integral_expr} can be rewritten as
\begin{equation} 
\Phi(q^kx)=\f{A}{2\pi i}\int_{C'} \exp\left(\f{1}{\epsilon}\left[\f{u^{\ell+2}}{\ell+2}-\sum_{j=0}^\ell \alpha_j u^j\right]\right)S_k(u) du\;,%
\label{eq_asymptotic_integral_expr_transformed}
\end{equation}
where $q=e^{-\epsilon}\to 1^-$, $C'$ is the image of the contour $C$ under $\op{T}^{-1}$ and $S_k(u) = \f{g(z(u))}{z(u)^k}\f{dz}{du}$. Now one writes 
\begin{equation}
S_k(u) = \sum_{j=0}^{\ell} p_j^{(k)} u^k + p'(u) H(u),
\label{eq_Gk_ansatz}
\end{equation}
where the $(p_j^{(k)})_{j=0}^\ell$ are analytic functions of the $(w_j)_{j=2}^{\ell}$ and $x$, and $H(u)$ is some analytic function of $u$. Substituting \cref{eq_Gk_ansatz} into \cref{eq_asymptotic_integral_expr_transformed}, one arrives at the asymptotic expression
\begin{align}
\Phi(q^kx)&=\label{eq_as_expr_phi}\\ &A\sum_{j=0}^{\ell} p_j^{(k)} \epsilon^{\f{j}{\ell+2}}\Theta_\ell^{(j)}(\alpha_1 \epsilon^{-\f{\ell+1}{\ell+2}},
\alpha_2\epsilon^{-\f{\ell}{\ell+2}},\dots,\alpha_{\ell}\epsilon^{-\f{2}{\ell+2}}),\nonumber
\end{align}
where $\Theta_\ell^{(0)}=\Theta_\ell$ and $\Theta_\ell^{(j)} =- \f{\partial}{\partial s_j} \Theta_\ell$ for $1\leq j \leq \ell$. By substituting \cref{eq_as_expr_phi} into \cref{eq_linearization_ansatz} for $k=0$ and $k=1$, we obtain an asymptotic expression for $G_{\ell}(w_2,\dots,w_\ell,x,q)$ which is valid uniformly with respect to the parameters $(w_j)_{j=2}^\ell$ and $x$, in particular in the vicinity of the multicritical point at which $\ell+1$ saddle points of the function $f(z)$ coalesce. Close to the multicritical point, the singular part of $G_{\ell}(w_2,\dots,w_\ell,x,q)$ satisfies the scaling relation
\begin{equation}
G^{\op{sing}}_{\ell}(w_2,\dots,w_\ell,x,e^{-\epsilon})=\epsilon^\theta F(\alpha_1 \epsilon^{-\phi_1},\dots,\alpha_\ell \epsilon^{-\phi_\ell}),
\end{equation}
where $F$ is expressible via $\Theta_\ell(s_1,\dots,s_\ell)$, $\theta=\f{1}{\ell+2}$, and the $\phi_j\equiv \phi_j(\ell)$ are given by \cref{eq_crit_exps}.

\end{document}